\documentstyle[aps]{revtex}
\input epsf         
\epsfverbosetrue    

\begin{document}
\twocolumn[\hsize\textwidth\columnwidth\hsize\csname@twocolumnfalse%
\endcsname

\title{Defect and anisotropic gap induced
quasi-one-dimensional modulation of local density of states in
YBa$_2$Cu$_3$O$_{7-\delta}$}

\author{Khee-Kyun Voo$^1$, Hong-Yi Chen$^2$, and W. C. Wu$^3$}
\address{$^1$Department of Physics, National Tsing-Hua
University, Hsinchu 30043, Taiwan\\ $^2$ Texas Center for
Superconductivity and Department of Physics, University of
Houston, Houston, TX 77204, USA\\ $^3$Department of Physics,
National Taiwan Normal University, Taipei 11650, Taiwan}

\date{November 30, 2002}
\maketitle \draft

\begin{abstract}
Motivated by recent angle-resolved photoemission spectroscopy
(ARPES) measurement that superconducting
YBa$_2$Cu$_3$O$_{7-\delta}$ (YBCO) exhibits a
$d_{x^2-y^2}$+$s$-symmetry gap, we show possible
quasi-one-dimensional modulations of local density of states in
YBCO. These aniostropic gap and defect induced stripe structures
are most conspicuous at higher biases and arise due to the nesting
effect associated with a Fermi liquid. Observation of these
spectra by scanning tunneling microscopy (STM) would unify the
picture among STM, ARPES, and inelastic neutron scattering for
YBCO.
\end{abstract}

\vspace{0.1cm}
]

After more than one and half decades of the discovery of
high-temperature superconductors, there lies still one central
question: are these materials behave like a Fermi liquid? Recent
scanning tunneling measurements (STM) however have provided a new
platform in detecting this critical issue, to which modulations of
local density of states (LDOS) are observed in superconducting
state of Bi$_2$Sr$_2$CaCu$_2$O$_8$ (BSCCO)
\cite{hoffman02-1,howald02,hoffman02-2}. More recently,
experimentilists managed to obtain Fourier transform (FT) of these
LDOS modulations, which then can be immediately used to test
whether the system is Fermi-liquid like.

Analogous to angular resolved photo-emission spectroscopy (ARPES),
FT-STM can study electronic properties connected to entire
Brillouin zone of the system. When a system behaves as a Fermi
liquid, Fermi surface (FS) scattering should be clearly identified
in normal state. While in superconducting state, it should also
give information on how the symmetry of Cooper pair order
parameter develops on the FS. Assuming the effect of defect,
experimentally observed STM LDOS modulation can be ascribed to
quantum interference arising because quasiparticles are scattered
by the defect locally. The occurrence of peaks and their evolution
as the bias change in the FT-STM spectra seem to be explained well
by Fermi liquid models\cite{tang02,wang02,zhang02}, whose FS
topology is consistent with previous ARPES results. Moreover, the
origin of peak moving in FT-STM is very similar to what of the
incommensurate peaks observed in inelastic neutron scattering
(INS) \cite{bourges00}. Apparently this scenario is in great
contrast to the stripe picture \cite{kivelson02}, interpreted in
the STM data of Howald {\em et al.} \cite{howald02} and the
one-dimensional (1D) INS data of Mook {\em et al.} \cite{mook00}.

Perhaps of technical involvement, the STM data done for YBCO so
far have been limited to the CuO chain layers only \cite{derro02}.
It becomes naturally to ask what would be seen by STM on CuO$_2$
plane layers of YBCO . As is well known, due to the coupling to
the 1D CuO chains, the CuO$_2$ plane is orthorhombic rather than
tetragonal. The perfect $D_{4h}$ symmetry is broken which accounts
naturally for the additional symmetry mixed in the order parameter
of the CuO$_2$ plane. In this regard, it is commonly accepted that
there is a subdominant $s$-wave component in addition to the
dominant $d_{x^2-y^2}$-wave superconducting gap for CuO$_2$
planes. This view is supported by earlier tunneling data
\cite{sun94} and verified by recent data of ARPES \cite{lu01}. The
1D like incommensurate INS peaks \cite{mook00} can also be
interpreted by this picture associated with a Fermi liquid
\cite{voo02}. When fitted to a simple $d_{x^2-y^2}$+$s$ model,
ARPES data reveals that the $s$-wave component is about 20 $\%$ of
$d$-wave component \cite{lu01}.

Here, based on a Fermi liquid model, we investigate the LDOS
specific to CuO$_2$ planes of superconducting YBCO. Similar to
other theoretical works \cite{tang02,wang02,zhang02}, we consider
the effect of defect, while pay special attention to the gap
anisotropy. Our goal is to bring out the most conspicuous features
and inspire STM measurements to be performed. Successful
observation of our predictions in either real-space or FT LDOS
could unambiguously identify (1) the goodness of Fermi-liquid
behavior and (2) the $d_{x^2-y^2}+s$ symmetry of gap for YBCO.
Furthermore this could lead to a unified picture among ARPES
\cite{lu01}, INS \cite{mook00}, and STM on YBCO.

To investigate the LDOS spectra, we start by considering the
following Hamiltonian

\begin{eqnarray} H=H_{\rm BCS}+H_{\rm I}, \label{eq:htot}
\end{eqnarray}
where the usual BCS part is

\begin{eqnarray}
H_{\rm BCS}=\sum_{{\bf k},\sigma}\xi_{\bf k} c^\dagger_{{\bf k}
\sigma}c_{{\bf k}\sigma}+\sum_{\bf k}\left[\Delta_{\bf k
}c^\dagger_{{\bf k}\uparrow}c^\dagger_{-{\bf k} \downarrow}+{\rm
H.c.}\right] \label{eq:hbcs}
\end{eqnarray}
and the part associated with the presence of one single extended
impurity located at site 0 is

\begin{eqnarray}
H_{\rm I}=&&\sum_{\langle i,j\rangle,\sigma} \delta t_{ij}
c^\dagger _{i\sigma}c_{j\sigma}+\sum_{\langle
i,j\rangle}\left[\delta\Delta_{ij}c^\dagger
_{i\uparrow}c^\dagger_{j\downarrow}+{\rm H.c.}\right]\nonumber\\
&+&V_0 \left(c^\dagger_{0\uparrow}c_{0\uparrow}+c^\dagger
_{0\downarrow}c_{0\downarrow}\right).
 \label{eq:hi}
\end{eqnarray}
The parameter $\delta t$ corresponds to local deviation of hopping
due to impurity , $\delta\Delta$ corresponds to local deviation of
pairing gap due to impurity, and $V_0$ is spin-independent
impurity potential. Any spin-dependent impurity potential ignoring
spin-flip will have no effect in the lowest-order approximation.
For extensive but weak impurity scattering, we consider local
deviations to those terms ($\delta t_1$, $\delta\Delta_1$) couple
the impurity site and its nearest neighbors and terms ($\delta
t_2$, $\delta\Delta_2$) couple impurity's nearest neighbors and
next nearest neighbors. Similar kinds of Hamiltonian have been
successfully used by Tang and Flatt\'{e} \cite{tang02} to explain
the resonant STM spectra for Ni impurities in BSCCO and by Wang
and Lee \cite{wang02} and Zhang and Ting \cite{zhang02} to explain
the energy-dependent modulation of FT-STM spectra on BSCCO. We
shall comment how our results are modified and their experimental
relevance when more than one defect are present later.

The local density of states measured by STM is given by

\begin{eqnarray}
D({\bf r},\omega)=&-&{1\over 2\pi}{\rm Im}[G_{11}({\bf r},{\bf
r},\omega+i0^+)\nonumber\\ &-&G_{22}({\bf r},{\bf
r},-\omega-i0^+)], \label{eq:dos}
\end{eqnarray}
where $G_{ij}$ is the element of the 2$\times$2 equal-location
single-particle Green's function matrix $\hat{G}$ in Nambu
representation. The full $\hat{G}$ can be calculated via the usual
Gor'kov-Dyson equation

\begin{eqnarray}
\hat{G}({\bf r},{\bf r},\omega)& =& \hat{G}^0({\bf r},{\bf
r},\omega)+\sum_{{\bf r}^\prime,{\bf
r}^{\prime\prime}}\hat{G}^0({\bf r},{\bf
r}^\prime,\omega)\hat{T}({\bf r}^\prime,{\bf
r}^{\prime\prime},\omega)\nonumber\\ &\times&\hat{G}^0({\bf
r}^{\prime\prime},{\bf r},\omega), \label{eq:ladder}
\end{eqnarray}
where $\hat{G}^0({\bf r}_i,{\bf r}_j)\equiv\hat{G}^0({\bf
r}_i-{\bf r}_j)$ correspond to the no-defect ones of $\hat{G}({\bf
r}_i,{\bf r}_j)$ and the sums (and the $T$-matrix $\hat{T}({\bf
r}^\prime,{\bf r}^{\prime\prime},\omega)$) in
Eq.~(\ref{eq:ladder}) are over the sites associated with $H_{\rm
I}$ in Eq.~(\ref{eq:hi}). For weak impurity scattering, we shall
calculate $\hat{G}$ in Eq.~(\ref{eq:ladder}) up to first order of
$\delta t$'s, $\delta\Delta$'s, and $V_0$, {\em i.e.}, in the Born
limit or first order $T$-matrix approximation \cite{zhang02}. In
our calculation, we use a $800\times 800$ square lattice sites and
the impurity is located in the center. For simple but reasonable
reason, we choose $2\delta t_1 = 4\delta
t_2=-2\delta\Delta_1=-4\delta\Delta_2=V_0$ and assume these scales
are samll such that the first order T-matrix approximation is
valid. It should be emphasized that the weak impurity scattering
limit is enough to bring out essential physics. Strong impurity is
not expected to give qualitatively new feature away from its
neighborhood of interest (see later). For $\xi_{\bf k}$, we use a
tight-binding band $\xi_{\bf k}=t_1({\rm cos}k_x+{\rm
cos}k_y)/2+t_2{\rm cos}k_x{\rm cos}k_y+t_3({\rm cos}2k_x+{\rm
cos}2k_y)/2 +t_4({\rm cos}2k_x{\rm cos}k_y+{\rm cos}k_x{\rm
cos}2k_y)/2+t_5{\rm cos}2k_x{\rm cos}2k_y$ (lattice constant
$a\equiv 1$), with $t_{1-5}=-0.60, 0.16, -0.05, -0.11, 0.05$ eV
respectively for YBCO. The fine structure of the band do not alter
the main results reported in this paper. For superconducting gap
of YBCO, we assume $\Delta_{\bf k}=\Delta_d/2(\cos k_x-\cos
k_y)+\Delta_s$ with $\Delta_d=36.7{\rm meV}$ and
$\Delta_s/\Delta_d\equiv s =0.2$. This gives the maximum gaps at
antinodes $(\pi,0)$ and $(0,\pi)$, $\Delta_{\rm max}^x=44{\rm
meV}$  and $\Delta_{\rm max}^y=29.3{\rm meV}$. Thus $\Delta_{\rm
max}^x/\Delta_{\rm max}^y=1.5$ in accordance with ARPES
measurement \cite{lu01}. In our calculation, we also introduce a
finite lifetime broadening $\gamma=2$ meV to the quasiparticle
Green's function to smooth our data points by replacing $\omega
+i0^+$ with $\omega +i\gamma$.

\begin{figure}[t]
\mbox{ \epsfxsize=1.25\hsize{\epsfbox{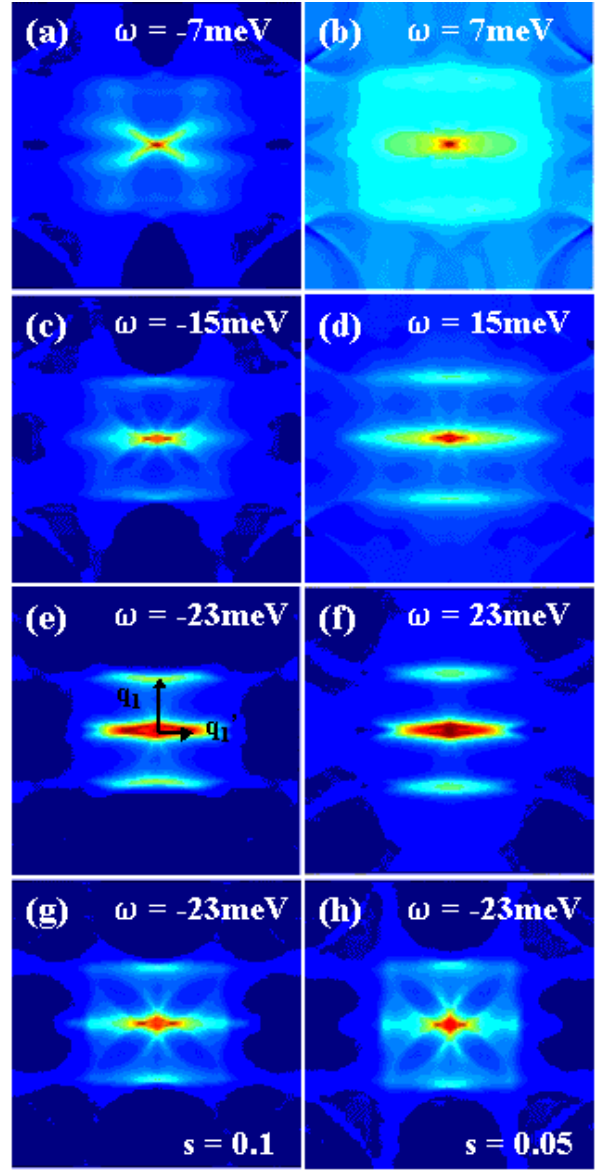}}} \vspace{0.1cm}
\caption{Bias energy dependent FT-LDOS plotted for the whole first
Brillouin zone for a $d_{x^2-y^2}+s$-wave superconductor. All
frames are for anisotropy $s=0.2$ except for (g) and (h) with
$s=0.1$ and $0.05$, which are intended to be compared to (e).
Wavevectors ${\bf q}_1$ and ${\bf q}_1^\prime$ in (e) are the
nesting ones shown in Fig.~\protect\ref{fig2}.} \label{fig1}
\end{figure}

Theoretically $D({\bf q},\omega)$, the Fourier transform of
$D({\bf r},\omega)$ is more readily obtained. In Fig.~\ref{fig1},
we calculate the bias energy dependent $D({\bf q},\omega)$ over
the whole first Brillouin zone. Energies have been taken from
$-23{\rm meV}$ to $23{\rm meV}$ --- a scale less than the maximum
gap at antinode $(0,\pi)$ of $\Delta_{\rm max}^y=29.3$ meV. All
frames are for anisotropy $s=0.2$ except (g) and (h) with $s=0.1$
and $0.05$, which are intended to be compared to (e). The case of
small $s$ recovers the results of Wang {\em et al.} \cite{wang02}
and Zhang {\em et al.} \cite{zhang02} for a pure
$d_{x^2-y^2}$-wave superconductor. When energy is low [case (a)
and (b)], only node-to-node scattering is allowed, to which the
finite ${\bf q}$ interference is not prominent. The most important
interference appears to those zero or small ${\bf q}$ scatterings,
which nevertheless exhibit clear feature of anisotropic gap. When
energy is high [case (e) and (f)], in contrast, finite ${\bf q}$
interference peaks emerge in addition to the strong interference
structure centered at ${\bf q}=0$.

\begin{figure}[t]
\mbox{\epsfxsize=0.75\hsize{\epsfbox{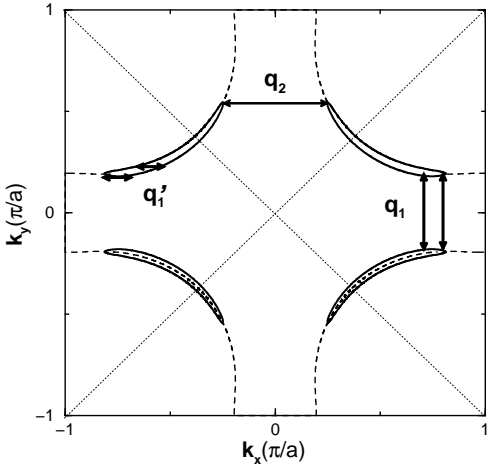}}} \vspace{0.5cm}
\caption{The Fermi surface (FS) used in our calculation (dash
lines) and contours of excitation energy $E_{\bf k}$ = 23 meV
(solid lines) are shown. Nesting wavevectors ${\bf q}_1$ and ${\bf
q}_1^\prime$ connecting locally almost parallel contours are
intimately related to interference peaks. While ${\bf q}_2$ is
shown as an example of no assocaited interference peak. For better
visualization, the FS within the upper contours are not plotted.}
\label{fig2}
\end{figure}

Fig.~\ref{fig2} shows the Fermi surface used in our calculation,
and contours of $E_{\bf k}$ = 23 meV. At this high energy (about
half of the maximum gap at antinode $(\pi,0)$, $\Delta_{\rm
max}^x=44$ meV), the 2-fold symmetry of the gap causes the contour
pair approaching ends of the bananas to be more parallel near
$x$-axis than the pairs near $y$-axis. The more parallel pair has
better nesting effect locally (intra- or inter-contour)
\cite{voo02}, which then further enhances the {\em joint} DOS . In
view of Fig.~1(e) or 1(f), conspicuous interference peaks
associated with wavevectors ${\bf q}_1$ and ${\bf q}_1^\prime$ are
clearly identified (see also Fig.~3). While, for example, no
similar interference peak associated with ${\bf q}_2$ (see
Fig.~\ref{fig2}) is seen.

\begin{figure}[t]
\mbox{ \epsfxsize=0.95\hsize{\epsfbox{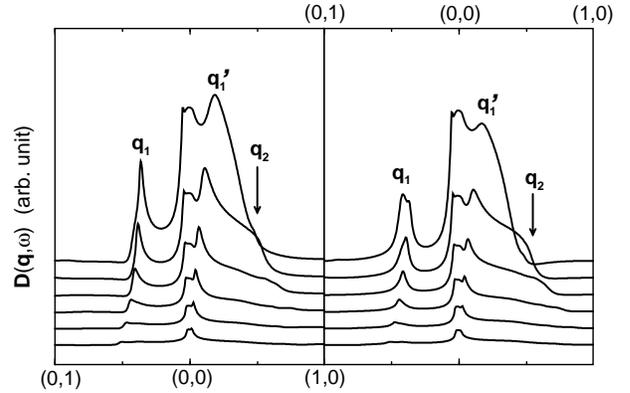}}} \vspace{0.5cm}
\caption{Scan of $D({\bf q},\omega)$ (arbitrary units) along ${\bf
q}a/\pi=$ (0,1)-(0,0)-(1,0) at bias energy from -5 to -23 meV
(left: from bottom to top), and from $+$5 to $+$23 meV (right:
from bottom to top) at equal energy intervals.} \label{fig3}
\end{figure}

To illustrate how the ${\bf q}_1$ and ${\bf q}_1^\prime$ peaks
evolve as the change of bias energy, we plot $D({\bf q},\omega)$
in Fig.~\ref{fig3}, taking ${\bf q}$ along ${\bf q}a/\pi=$
(0,1)-(0,0)-(1,0) at bias energy from -5 to -23 meV (left: from
bottom to top), and from $+$5 to $+$23 meV (right: from bottom to
top). Both frames show clear signatures of ${\bf q}_1$ and ${\bf
q}_1^\prime$ peaks which disperse as the bias energy changes.
Their intensity increases as bias energy increases, to which ${\bf
q}_1$ disperses convergently to a fixed value near $\pm
0.25(2\pi/a)~\hat{\bf y}$, while ${\bf q}_1^\prime$ disperses
divergently. No peak develops at ${\bf q}_2$ except some plateaus
at intermediate energies. The extensive, centered at zone center
$(0,0)$, quasi-1D like strong interference structure (parallel to
$k_x$ axis, see Fig.~\ref{fig3} as well as Fig.~\ref{fig1}(e)),
comprised of ${\bf q}_1^\prime$ peaks discussed above, is
superposed by the zero and small-${\bf q}$ scattering connecting
two ${\bf k}$ points within the same contour. It is the ${\bf
q}_1^\prime$ peaks, which is enhanced by local nesting effect,
gives rise to the 1D interference structure parallel to $k_x$
axis.

One particular interest on $D({\bf q},\omega)$ is the {\em two}
${\bf q}_1$ peaks which have wavevectors near $\pm
0.25(2\pi/a)~\hat{\bf y}$. These two peaks, if by themselves
alone, would imply a 1D parallel to $x$-axis stripe pattern in the
real-space DOS (with a period of roughly $4a$) --- a pattern
similar to the phenomenon predicted by ``stripe'' model
\cite{kivelson02}.

\begin{figure}[h] \mbox{
\epsfxsize=0.92\hsize{\epsfbox{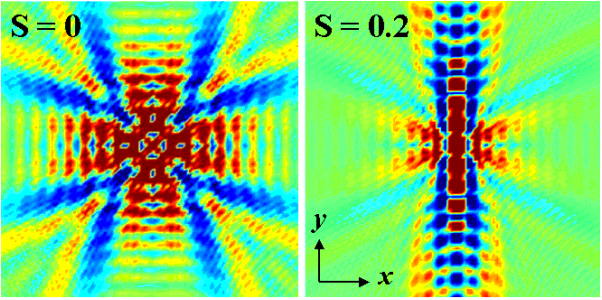}}} \vspace{-6.75cm}
\caption{Modulations of real-space LDOS for bias frequency
$\omega=-23$ meV. The left (right) frame is for a
$d_{x^2-y^2}$+$s$-wave superconductor with $s=0$ (0.2) and both
are presented for $x,y=-40$ to $40a$. Single impurity is located
at the center.} \label{fig4}
\end{figure}

To see how these stripes develop in real-space, in Fig.~\ref{fig4}
we present the real-space LDOS modulation, taking Fourier
transform of $D({\bf q},\omega)$.  We consider only the
$\omega=-23$ meV and  $s=0.2$ case [Fig.~1(e)] and include the
$s=0$ case for comparison. For $s=0.2$ case, quasi-1D chains of
beads of roughly $4a$ wide oriented approximately along $x$ axis
are clearly seen at places away from the impurity site. The
occurrence of these Friedel-type oscillating stripes arises
because of the strong, centered at $(0,0)$, 1D interference
structures. The latter also leads to the variation of intensities
along $y$ axis. In contrast, for the four-fold symmetry $s=0$
case, similar 1D stripes also develop at places away from the
impurity site. Clearly there exists no cross stripe pattern
(checkerboard) anywhere in the $s=0$ case, in contradiction to
most theoretical predictions \cite{wang02,zhang02}. Recently
Zhang, Hu, and Yu \cite{zhanggm02} considered the Zn or Ni
impurity induced modulations of real-space LDOS for a
$d_{x^2-y^2}$-wave superconductor which has a simple continuum
band of circular FS. They \cite{zhanggm02} reported similar
circularly one-dimensional (not cross) patterns.

It should be emphasized that all the experimental ${\bf q}$-space
peaks are obtained upon Fourier transforming a single chosen piece
of real space, which shows no sign of localized impurities. While
in our approach, it is clear that the different ${\bf q}$ peaks
are originated from different patches of space surrounding the
impurity (see Fig.~4). To relate our calculation ($s=0$ case) to
the experiments (checkerboard pattern), we need to have {\em
dilute} impurities, such that the decaying Friedel oscillations
emanating from individual impurities overlap crossly at some open
spaces, and therefore Fourier transforming this single region
gives all the ${\bf q}$ peaks.

A direct consequence of this picture is that when the impurity
concentration is too high, the LDOS oscillations will be smoothed
out, as the phases of the oscillations are pinned by individual
impurities and therefore {\em incoherent}. This will be in
contrast to the genuine stripe scenario \cite{kivelson02} where
the oscillation will be expected to be more prominent since the
stripes, fluctuating in smooth real spaces, will now get pinned by
the impurities.

The case of multiple impurities for BSCCO is recently considered
by Morr {\em et al.} \cite{morr02} and Zhu {\em et al.}
\cite{zhu02}. Very recently Anderson and Hedeg{\aa}rd
\cite{anderson03} also study the quantum interference between
multiple impurities for a $d_{x^2-y^2}$-wave superconductor, whose
pairing state could become $d_{x^2-y^2}+id_{xy}$ or
$d_{x^2-y^2}+is$-wave at some doping level. Our dilute
multi-impurity regime discussed above is fundamentally different
from that any one else \cite{morr02,zhu02,anderson03} has studied.
Let the length scale between the impurities be $d_0$, and the
distance away from the impurities, where we are interested, be
$d$. Our interesting regime is then $d_0 \sim d\agt 20a$ (see
Fig.~4), while the case other people have studied was $d_0\sim
d\sim$ a few $a$.

Finally one does not expect any qualitatively new physics at the
use of an unitary impurity, since it is known to affect only the
LDOS locally close to the impurities, whereas we are now most
interested in the regions away from the impurities. The essential
point is that impurity scatters the quasiparticles, no matter in
Born or unitary limit.

In summary, motivated by recent ARPES and INS results which are
consistent with the picture that YBCO has superconducting gap of
$d_{x^2-y^2}$+$s$ symmetry,  we predict that quasi-one-dimensional
Friedel-type stripes can be induced by defect in superconducting
YBCO, observable by STM measurement. This phenomenon is most
conspicuous at higher biases and associated with the local nesting
effect for a Fermi liquid. Hopefully the predicted modulations in
either real or Fourier space can be tested by STM measurement
soon.

This work is supported by National Science Council of Taiwan under
Grant No. 91-2112-M-003-020.


\begin{references}

\bibitem{hoffman02-1}
{J.E. Hoffman, E.W. Hudson, K.M. Lang, V. Madhavan, H. Eisaki, S.
Uchida, J.C.
  Davis}, Science {\bf 295},  466  (2002).

\bibitem{howald02}
C. Howald, H. Eisaki, N. Kaneko, and A. Kapitulnik,
cond-mat/0201546.

\bibitem{hoffman02-2}
{J.E. Hoffman, K. McElroy, D.H. Lee, K.M. Lang, H. Eisaki, S.
Uchida, J.C.
  Davis}, Science {\bf 297},  1148  (2002).

\bibitem{tang02}
{J.-M. Tang and M.E. Flatt\'{e}}, Phys. Rev. B {\bf 66},  060504
(2002).

\bibitem{wang02}
Q.-H. Wang and D.H. Lee, cond-mat/0205118.

\bibitem{zhang02}
D. Zhang and C.S. Ting, cond-mat/0209318.

\bibitem{bourges00}
{P. Bourges, Y. Sidis, H.F. Fong, L.P. Regnault, J. Bossy, A.
Ivanov, and B.
  Keimer}, Science {\bf 288},  1234  (2000).

\bibitem{kivelson02}
S.A. Kivelson {\em et al.}, cond-mat/0210683.

\bibitem{mook00}
{H.A. Mook, P. Dai, F. Dogan, and R.D. Hunt}, Nature {\bf 404},
729  (2000).

\bibitem{derro02}
{D.J. Derro, E.W. Hudson, K.M. Lang, S.H. Pan, J.C. Davis, J.T.
Markert, and
  A.L. de Lozanne}, Phys. Rev. Lett. {\bf 88},  097002  (2002).

\bibitem{sun94}
{A.G. Sun, D.A. Gajewski, M.B. Maple, and R.C. Dynes}, Phys. Rev.
Lett. {\bf
  72},  2267  (1994).

\bibitem{lu01}
{D.H. Lu, D.L. Feng, N.P. Armitage, K.M. Shen, A. Damascelli, C.
Kim, F.
  Ronning, Z.-X. Shen, D.A. Bonn, R. Liang, W.N. Hardy, A.I. Rykov, and S.
  Tajima}, Phys. Rev. Lett. {\bf 86},  4370  (2001).

\bibitem{voo02}
{K.K. Voo, H.Y. Chen, and W.C. Wu}, Physica C {\bf 382},  323
(2002).

\bibitem{zhanggm02}
{G.-M. Zhang, H. Hu, and L. Yu}, Phys. Rev. B {\bf 66},  104511
(2002).

\bibitem{morr02}
{D.K. Morr and N.A. Stavropoulos}, Phys. Rev. B {\bf 66},
140508(R)  (2002).

\bibitem{zhu02}
L. Zhu, W.A. Atkinson, and P.J. Hirschfeld, cond-mat/0208008.

\bibitem{anderson03}
B.M. Anderson and P. Hedeg{\aa}rd, cond-mat/0301225.

\end{references}
\end{document}